\documentclass[aps, twocolumn, superscriptaddress,amsmath,amssymb,aps,
pra]{revtex4-1}

\pdfoutput=1

\usepackage{braket}             
\usepackage{graphicx}           
\usepackage{bm}             
\usepackage[usenames, dvipsnames]{color}
\usepackage{hyperref}           
\usepackage{mathtools}
\usepackage[normalem]{ulem} 
\usepackage[caption=false]{subfig}
\usepackage{float}
\usepackage{graphicx}

\usepackage{soul}

\usepackage{colonequals} 

\usepackage[usenames, dvipsnames]{xcolor}

\begin{document}
\title{Reservoir engineering strong quantum entanglement in cavity magnomechanical systems}

\author{Zhi-Qiang Liu}
\affiliation{Lanzhou Center for Theoretical Physics, Key Laboratory of Theoretical Physics of Gansu Province, Lanzhou University, Lanzhou, Gansu $730000$, China}
\author{Yun Liu}
\affiliation{Lanzhou Center for Theoretical Physics, Key Laboratory of Theoretical Physics of Gansu Province, Lanzhou University, Lanzhou, Gansu $730000$, China}
\author{Lei Tan}
\email{tanlei@lzu.edu.cn}
\affiliation{Lanzhou Center for Theoretical Physics, Key Laboratory of Theoretical Physics of Gansu Province, Lanzhou University, Lanzhou, Gansu $730000$, China}
\affiliation{Key Laboratory for Magnetism and Magnetic Materials of the Ministry of Education, Lanzhou University, Lanzhou $730000$, China}
\author{Wu-Ming Liu}
\affiliation{Beijing National Laboratory for Condensed Matter Physics, Institute of Physics, Chinese Academy of Sciences, Beijing 100190, China}

\date{\today}
\begin{abstract}
We construct a hybrid cavity magnomechanical system to transfer the bipartite entanglements and achieve the strong microwave photon-phonon entanglement based on the reservoir engineering approach. The magnon mode is coupled to the microwave cavity mode via magnetic dipole interaction, and to the phonon mode via magnetostrictive force (optomechanical-like). It is shown that the initial magnon-phonon entanglement can be transferred to the photon-phonon subspace in the case of these two interactions cooperating. In the reservoir-engineering parameter regime, the initial entanglement is directionally transferred to the photon-phonon subsystem, so we obtain a strong bipartite entanglement in which the magnon mode acts as the cold reservoir to effectively cool the Bogoliubov mode delocalized over the cavity and the mechanical deformation mode. Moreover, as the dissipation ratio between the cold reservoir mode and the target mode increases, greater quantum entanglement and better cooling effect can be achieved. Our results indicate that the steady-state entanglement is robust against temperature. The scheme may provide potential applications for quantum information processing, and is expected to be extended to other three-mode systems.
\end{abstract}


\maketitle

\section{Introduction}
In recent years, ferromagnetic systems as a new research field has attracted extensive attention. In particular, yttrium iron garnet (YIG) materials exhibit excellent ferromagnetic properties due to their high Curie temperature and large spin density~\cite{Huebl:2013,Tabuchi:2014,Goryachev:2014,Zhang:2014}. In YIG spheres, the collective excitation of a large number of spins called magnons, which can be strongly coupled to microwave photons and superconducting qubits theoretically~\cite{Blais:2004,Bai:2015,Bourhill:2016} and experimentally~\cite{Kostylev:2016,Wallraff:2004,Tabuchi:2015,ZhangD:2015}. The YIG sphere is represented as a magnon mode and a mechanical vibration mode when it is driven by an external magnetic field. Note that the changing magnetization induced by the magnon excitation inside the YIG sphere leads to the deformation of its geometric configuration, which forms the phonon mode~\cite{Lachance-Quirion:2019}. In a standard cavity magnomechanical system consisting of a microwave cavity and a YIG sphere, the magnon couples to the microwave photon with magnetic dipole interaction and to the phonon via magnetostrictive interaction~\cite{ChenYT:2021}. Based on the nonlinear radiation pressure-like interaction between the deformation and magnetostatic modes, many physical phenomena originally developed in cavity optomechanical can be applied to cavity magnomechanical, such as magnomechanically induced transparency~\cite{WangB:2018,Gevorgyan:2022}, the observation of bistability~\cite{WangYP:2018}, magnetic nonreciprocity~\cite{WangYP:2019,Zhang:2020,Zhao:2020}, and high-order sidebands preparation~\cite{WangM:2021}. Particularly, the realization of quantum entanglement in hybrid multimode cavity-magnetic systems has attracted wide interest. In a cavity magnomechanical system, the genuine magnon-photon-phonon tripartite entanglement has been achieved with the assistance of nonlinear magnetostrictive interactions~\cite{Li:2018}. Later, it is shown that the magnon-magnon entanglement can be generated by magnomechanical interaction~\cite{Li:2019,Nair:2020}. Zhang $et$ $al$. placed two YIG spheres in a microwave cavity and generated entanglement between two magnon modes via Kerr nonlinear effects~\cite{ZhangZ:2019}. The subsequent scheme shows that the macroscopic entanglement between two YIG spheres can be achieved by injecting a two-mode squeezed vacuum field even without any nonlinear interaction~\cite{Yu:2020}. However, the steady-state entanglement created by these ideas is small, which is detrimental to the experimental observation of entanglement and its practical application.

Moreover, many cavity quantum electrodynamics (QED) systems is normally unavoidable interaction with an external environment noise which can lead to decoherence for system, where the system loses entanglement in finite time~\cite{Farace:2012,Mari:2009,Hu:2019,Mari:2012,WangG:2014,Hu:2020}. Thus, it is necessary to obtain and observe large order of magnitude entanglement for its practical application. Wang $et$ $al$. consider a three-mode bosonic coupled optomechanical system consisting of one mechanical mode and two optical modes, where the mechanical mode as a cold reservoir effectively cooling a delocalized Bogoliubov mode of the other modes, which is called the reservoir engineering approach~\cite{WangYD:2013}. It is shown that large steady-state cavity-cavity entanglement can be achieved when the mechanical mode is far more dissipative than the two cavity modes. Another scheme finds strong and stationary entanglement between the mechanical mode and the cavity field by using the other intermediate mechanical mode as a cold dissipative reservoir and by combining Coulomb interaction with optomechanical coupling~\cite{Chen:2015}. Hussain $et$ $al$. showed that bipartite and tripartite entanglement also can be enhanced by adding an optical parametric amplifier to the standard cavity magnomechanical system~\cite{Hussain:2022}. However, the stationary entanglement achievable via this reservoir engineering approach can exceed the limit of entanglement obtained by the coherent two-mode squeezing interaction due to the stability constraints of the continuous variable (CV) system~\cite{Vitali:2007,Yin:2009,Yin:2019}. In addition, reservoir engineering is independent of the initial state of the system and is robust to decoherence, which enables one to create and verify entanglement experimentally~\cite{Krauter:2011,Kronwald:2014,Kraus:2004,Liu:2021,Woolley:2014}. The scheme of obtaining quantum entanglement in hybrid cavity magnomechanical systems via reservoir engineering is still lacking. So, we are working on the directional transfer and enhancement of entanglement, which leads us to get the strong two-mode entanglement we need in multimode systems.

\begin{figure}[!htbp]
\centering
{\includegraphics[width=1.00\linewidth]{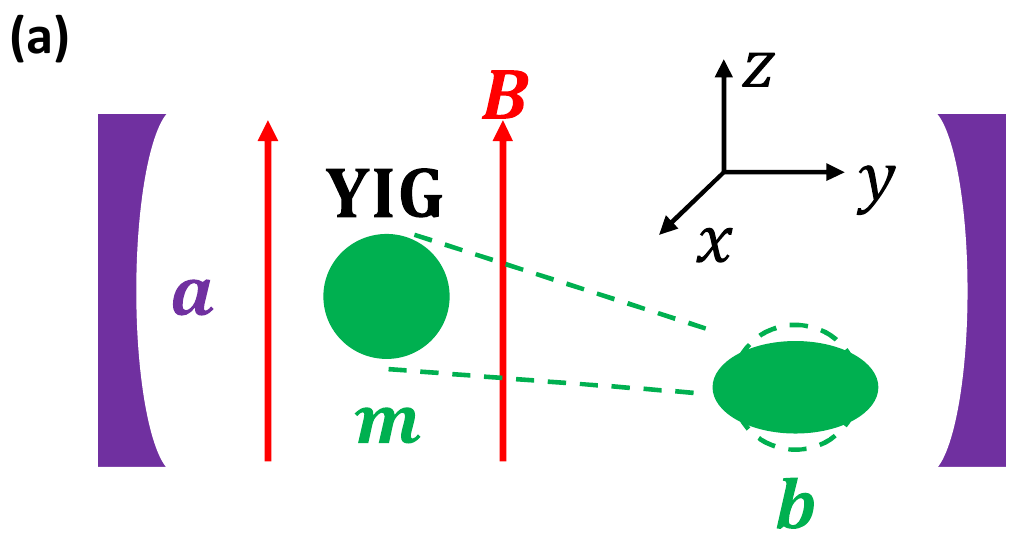}}\\
{\includegraphics[width=1.00\linewidth]{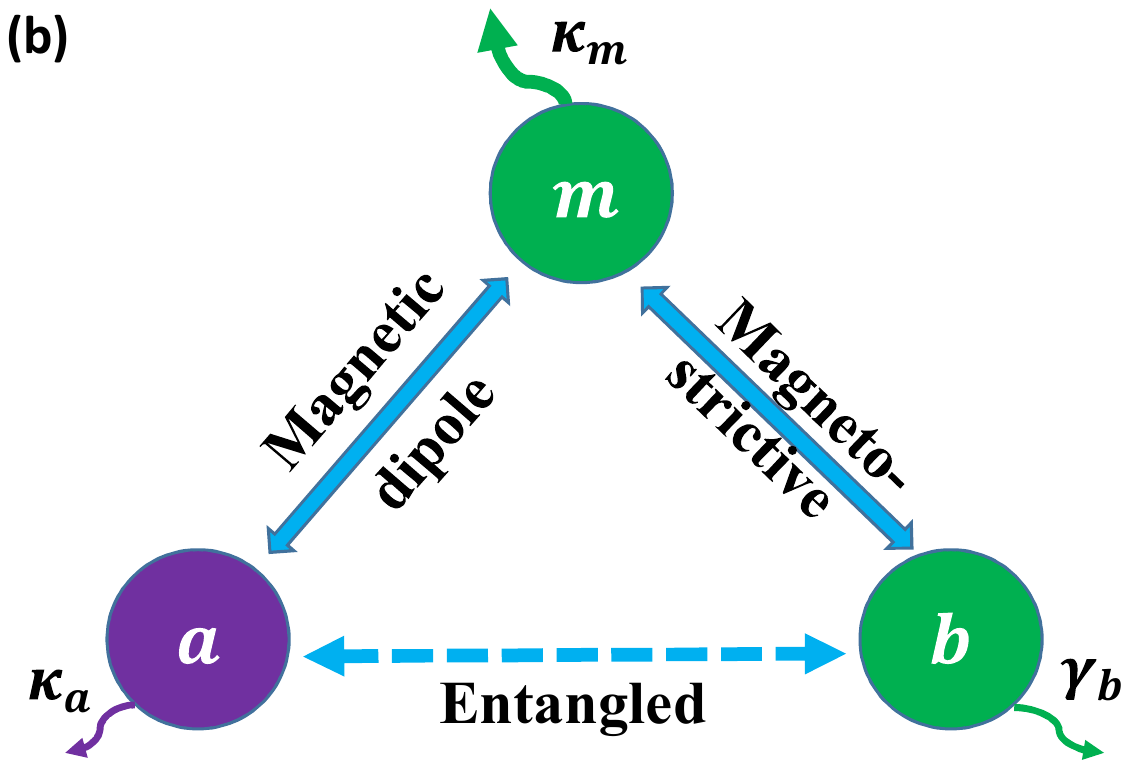}}\qquad
\caption{(Color online) (a) Schematic diagram of a standard cavity magnomechanical setup. The YIG sphere is placed inside the cavity near the maximum magnetic field of the cavity mode $a$ along the $x$ axis, and simultaneously in an uniform bias magnetic field of intensity $B$ along the $z$ axis, which together promotes the coupling between the microwave photon and the magnon. Furthermore, the YIG is driven directly by an external magnetic field (not shown) along the $y$ axis to generate a phonon mode, which is coupled with the magnon mode of the YIG via magnomechanical interaction. (b) Sketch of the theoretical model of the system. The magnon mode $m$ couples to the photon mode $a$ by the magnetic dipole, while $m$ and the phonon mode $b$ are coupled via magnetostrictive interaction. The mode $m$ has a greater dissipation rate than the other modes, so that it can act as a cold reservoir to engineering the photon-phonon entanglement.}\label{mag}
\end{figure}

In this paper, we construct a standard cavity magnomechanical setup including a microwave cavity $a$ and a YIG sphere as shown in Fig. \ref{mag}$(a)$. The deformation mode induced by the YIG sphere driven by an extra magnetic field is coupled with the magnetostatic mode of the YIG through magnetostrictive forces, and the latter simultaneously interacts with the microwave cavity mode via a magnetic dipole, while there is no direct interaction between the cavity mode and the mechanical mode as shown in Fig. \ref{mag}$(b)$. We find that all two-mode entanglements of the system arise from the radiation pressure-like interaction, and that the initial entanglement can be transferred between different subsystems through the beam-splitter interaction, which enables one to generate the photon-phonon entanglement when the system is stable. We focus on the reservoir engineering parameter region that allows for effectively cooling the Bogoliubov mode hybridized over the cavity mode and the deformation mode via the cold reservoir mode, thus the large steady-state photon-phonon entanglement can be generated that surpass the limit of entanglement created by the coherent parametric interaction. The stationary entanglement between the photon mode and the phonon mode is robust to environmental temperature. The reservoir engineering scheme is also possible in other three-mode quantum systems.

The paper is organized as follows: In Sec. \uppercase\expandafter{\romannumeral2}, we introduce our physical model and derive the linearized Hamiltonian of the system via quantum Langevin equations and standard linearization techniques. Sec. \uppercase\expandafter{\romannumeral3} is devoted to obtaining the stationary photon-phonon entanglement and analyzing its mechanism. In Sec. \uppercase\expandafter{\romannumeral4}, we discuss the reservoir-engineered strong steady-state entanglement between the deformation mode and the microwave cavity mode and show the directional transfer of entanglement. Finally, the conclusions are summarized in Sec. \uppercase\expandafter{\romannumeral5}.

\section{MODEL AND HAMILTONIAN}
We consider a standard cavity magnomechanical system involving an uniform magnetostatic mode $m$, a microwave cavity $a$ and a deformation mode $b$ induced by the magnon excitation inside the YIG sphere, as shown in Fig. \ref{mag}. The magnetostatic mode and the cavity mode are coupled with each other via magnetic dipole interaction, in cooperation with the magnetic field of the cavity mode and the bias magnetic field. The deformation mode couples to the magnetostatic mode of the YIG via nonlinear magnetostrictive effect induced by a strong microwave driving field with frequency $\omega_l$ and amplitude $B_1$. The Hamiltonian of the system is given by ($\hbar=1$)
\begin{align}
H=&\omega_{c}a^{\dagger}a+\omega_{m}m^{\dagger}m+\omega_{b}b^{\dagger}b+g_{bm}m^{\dagger}m(b+b^{\dagger}) \nonumber\\
&+g_{am}(a+a^{\dagger})(m+m^{\dagger})+i\Omega (m^{\dagger}e^{-i \omega_{l}t}-me^{i \omega_{l}t}),\label{H1}
\end{align}
where $a$ ($a^{\dagger}$), $m$ ($m^{\dagger}$), and $b$ ($b^{\dagger}$) are the annihilation (creation) operators of the cavity, magnon, and mechanical modes, respectively, and satisfy the standard bosonic commutation relation; and $\omega_c$, $\omega_m$, and $\omega_b$ are the resonance frequency of the photon, magnon, and phonon modes, respectively. Particularly, the circular frequency $\omega_m$ is tunable by the bias magnetic field $B$ and the gyromagnetic ratio $\gamma_y$ of the YIG, i.e., $\omega_m=\gamma_yB$. The $g_{bm}m^{\dagger}m(b+b^{\dagger})$ term represents the interaction between the magnon mode and the mechanical membrane induced by the radiation pressure-like with the single-magnon coupling strength $g_{bm}$. Similar to optomechanical coupling, this nonlinear magnomechanical interaction is weak in nature, however it can be intensified to the strongly coupled parameter region by directly driving the YIG with a strong microwave field~\cite{WangYP:2016}. The microwave photon-magnon coupling rate $g_{am}$ includes a beam-splitter interaction and a parametric amplification, and it enters into strong coupling regime (i.e., $g_{am}/\kappa_a,\kappa_m>1$) as has been reported in Refs.~\cite{Huebl:2013,Goryachev:2014}. The last term in Eq. (\ref{H1}) describes the coupling between the pumping field $B_1$ and the magnon mode with coupling rate $\Omega=\frac{\sqrt{5}}{4}\gamma_y\sqrt{N}B_1$, in which the total number of spins is defined as $N=\rho V$ with the spin density $\rho=4.22\times10^{27}m^{-3}$ and the volume $V$ of the YIG sphere~\cite{ZhangX:2016}. Note that in the Hamiltonian of the system we have omitted the radiation pressure term because the size of the sphere (250-$\mu$m-diameter) chosen is much smaller than the microwave wavelength.

Then the Hamiltonian (\ref{H1}), in the rotating frame with respect to the laser driving frequency $\omega_l$, can be written as
\begin{align}
H^{\prime}=&\Delta_{c}a^{\dagger}a+\Delta_{m}^{\prime}m^{\dagger}m+\omega_{b}b^{\dagger}b+g_{bm}m^{\dagger}m(b+b^{\dagger}) \nonumber\\
  &+g_{am}(a^{\dagger}m+am^{\dagger})+i\Omega (m^{\dagger}-m),\label{Hrf}
\end{align}
where $\Delta_{c}=\omega_{c}-\omega_{l}$ and $\Delta_{m}^{\prime}=\omega_{m}-\omega_{l}$ are the detunings of the microwave cavity and magnon mode frequencies to the driving laser frequency, respectively. We neglect the anti-rotating-wave approximation term in Hamiltonian (\ref{Hrf}) when the condition $|2\omega_{l}| \gg g_{am}$ is satisfied. Considering the dissipation and input noises, the quantum Langevin equations (QLEs) governing the dynamical behavior of the system are given by
\begin{align}
\dot a =&-i\Delta_{c}a-ig_{am}m-\frac{\kappa_{a}}{2}a+\sqrt{\kappa_{a}}a_{in}, \nonumber\\
\dot m =&-i\Delta_{m}^{\prime}m-ig_{am}a-ig_{bm}m(b+b^{\dagger})+\Omega\nonumber\\
&-\frac{\kappa_{m}}{2}m+\sqrt{\kappa_{m}}m_{in}, \nonumber\\
\dot b =&-i\omega_{b}b-ig_{bm}m^{\dagger}m-\frac{\gamma_{b}}{2}b+\sqrt{\gamma_{b}}b_{in},\label{QLEs}
\end{align}
here $\kappa_a$ and $\kappa_m$ are the dissipation rates of the microwave cavity and magnon mode, respectively, and $\gamma_b$ is the mechanical damping rate. $a_{in}$, $m_{in}$, and $b_{in}$ are input noise operators for the microwave, mechanical, and magnon modes, respectively, with zero-mean values and whose nonzero correlation functions under Markovian approximation are $\langle o_{in}(t)o_{in}^{\dagger}(t^{\prime})\rangle=(\overline{n}_{o}+1)\delta(t-t^{\prime})$ and $\langle o_{in}^{\dagger}(t)o_{in}(t^{\prime})\rangle=\overline{n}_{o}\delta(t-t^{\prime})$, where $\overline{n}_{o}=[\exp(\hbar\omega_{o}/k_{B}T)-1]^{-1}$ ($o=a, m, b$) are the mean thermal photon, magnon, and phonon number, respectively, and $k_B$ is the Boltzmann constant. Note that the magnon mode as a bosonic mode satisfying the bosonic commutation rules must be at low excitations according to the Holstein-Primakoff transformation~\cite{Kusminskiy:2019}, i.e. $\langle m^{\dagger}m\rangle \ll 2Ns$, where $s=\frac{5}{2}$ is the spin number of the ground state $Fe^{3+}$ ion in YIG.

In the case of strong coherent driving, one can rewrite the Heisenberg operators as $a=\alpha+\delta a$, $m=\epsilon+\delta m$, and $b=\beta+\delta b$, where $\delta a$, $\delta m$, and $\delta b$ are the quantum fluctuation operators with zero mean values around the classical steady-state amplitudes $\alpha$, $\epsilon$, and $\beta$, respectively, and $|\alpha|, |\epsilon|, |\beta| \gg 1$. Thus, the standard linearization techniques can be applied to Eq. (\ref{QLEs}), we then obtain
\begin{align}
\dot {\alpha} &=-i\Delta_{c}\alpha-ig_{am}\epsilon-\frac{\kappa_{a}}{2}\alpha,\nonumber \\
\dot {\epsilon} &=-i\Delta_{m}\epsilon-ig_{am}\alpha+\Omega-\frac{\kappa_{m}}{2}\epsilon,\nonumber \\
\dot {\beta} &=-i\omega_{b}\beta-ig_{bm}|\epsilon|^{2}-\frac{\gamma_{b}}{2}\beta,\label{classicalEs}
\end{align}
where $\Delta_{m}=\Delta_{m}^{\prime}+g_{bm}(\beta+\beta^{*})$ is the effective detuning of the magnog mode to the YIG driving laser, which is modified by the mechanical vibration. The classical steady-state solution of the system can be obtained in the long-time limit
\begin{align}
\alpha &=\frac{-ig_{am}\epsilon}{i\Delta_{c}+\kappa_{a}/2},\nonumber \\
\epsilon &=\frac{-ig_{am}\alpha+\Omega}{i\Delta_{m}+\kappa_{m}/2},\nonumber \\
\beta &=\frac{-ig_{bm}|\epsilon|^{2}}{i\omega_{b}+\gamma_{b}/2}.\label{classical values}
\end{align}
Simultaneously, by neglecting the higher-order nonlinear terms, the linearized QLEs describing the quantum fluctuation operators can be written as
\begin{align}
\dot {\delta a} =&-i\Delta_{c}\delta a-ig_{am}\delta m-\frac{\kappa_{a}}{2}\delta a+\sqrt{\kappa_{a}}a_{in},\nonumber \\
\dot {\delta m} =&-i\Delta_{m}\delta m-ig_{am}\delta a-ig_{bm}\epsilon(\delta b+\delta b^{\dagger})\nonumber \\
&-\frac{\kappa_{m}}{2}\delta m+\sqrt{\kappa_{m}}m_{in},\nonumber \\
\dot {\delta b} =&-i\omega_{b}\delta b-ig_{bm}(\epsilon^{*}\delta m+\epsilon\delta m^{\dagger})-\frac{\gamma_{b}}{2}\delta b+\sqrt{\gamma_{b}}b_{in}.\label{quantumEs}
\end{align}

According to Eq. (\ref{quantumEs}), we can naturally obtain the linearized Hamiltonian of the system for the quantum fluctuation operators
\begin{align}
H_{lin}=&\Delta_{c}\delta a^{\dagger}\delta a+\Delta_{m}\delta m^{\dagger}\delta m+\omega_{b}\delta b^{\dagger}\delta b+g_{am}(\delta m^{\dagger}\delta a\nonumber\\
&+\delta m\delta a^{\dagger})
+[G_{bm}(\delta m^{\dagger}\delta b+\delta m\delta b)+H.c.],\label{Hlin}
\end{align}
where $G_{bm}=g_{bm}\epsilon$ is the enhanced magnomechanical coupling rate and assumed to be real for simplicity. Since the microwave cavity mode $\delta a$ has almost the same frequency as the magnetostatic mode $\delta m$, i.e., $\omega_{c}\approx \omega_{m}$, we first set $\Delta_{c}=\Delta_{m}$. Then, we pump the magnon mode with the driving at the blue sideband with respect to the mechanical mode, i.e., $\Delta_{m}=-\omega_{b}$. Furthermore, we assume that the optomechanical-like interaction is much smaller than the frequency of the mechanical deformation mode $\delta b$, $G_{bm} \ll 2\omega_{b}$. Therefore, we can obtain the resonant interaction Hamiltonian under the rotating-wave approximation (RWA) condition as follows
\begin{align}
H_{int}=g_{am}(\delta m^{\dagger}\delta a+\delta m\delta a^{\dagger})+G_{bm}(\delta m\delta b+\delta m^{\dagger}\delta b^{\dagger}),\label{Hlrint}
\end{align}
where we have constructed the desired linearized interaction term in a three-mode cavity magnomechanical system, which satisfies the mechanism of reservoir engineering as in cavity optomechanical systems. Next we will show by comparison that the strong steady-state entanglement and the directional transfer of entanglement can be realized by using the reservoir engineering approach.

\section{Steady-state entanglement and its transfer}
In this section, we focus on the steady-state entanglement of the system and the mechanism that produces the CV bipartite entanglement. The experimentally reachable parameters are $\omega_{c}/2\pi=10GHz$, $\omega_{b}/2\pi=10MHz$, $G_{bm}/2\pi=3.2MHz$, $g_{am}/2\pi=3.2MHz$, $\kappa_{m}/2\pi=\kappa_{a}/2\pi=1MHz$, $\gamma_{b}/2\pi=10^{2}Hz$, and the driving magnetic field $B_{1}\simeq3.9\times10^{-5}T$ corresponding to the driving power $P=8.9mW$~\cite{Li:2018,ZhangX:2016}. First, we need to analyze the stationary dynamics and find the stable parameter region of the system before quantifying the entanglement. Since the quantum noise source is Gaussian and the interaction between three modes is linearized, it ensures that our system is always in a Gaussian state. For CV systems, the Routh-Hurwitz criterion~\cite{DeJesus:1987} tells us that the system will be stable when all eigenvalues of the coefficient matrix $A$ have negative real parts. Then, one can calculate the bipartite entanglement of any pair of modes by the logarithmic negativity $E_{N}$, which is derived in detail in the Appendix.

\begin{figure}[!htbp]
\centering
{\includegraphics[width=1.00\linewidth]{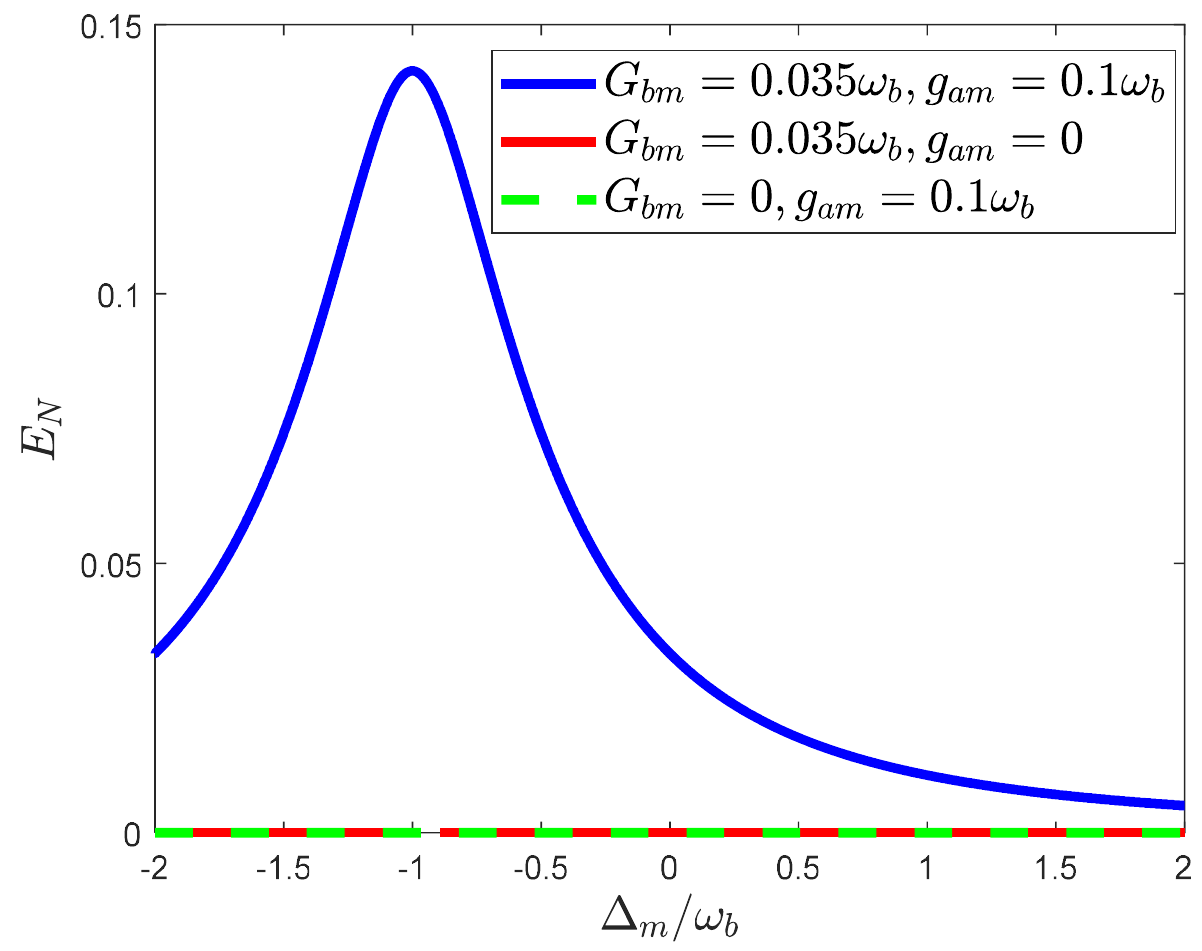}}\qquad
\caption{(Color online) The steady-state microwave photon-phonon entanglement $E_N$ versus the effective detuning $\Delta_{m}/\omega_b$ of the magnon mode frequency to the driving laser frequency for $G_{bm}=0.035\omega_b$, $g_{am}=0.1\omega_b$ (blue solid line), $G_{bm}=0.035\omega_b$, $g_{am}=0$ (red solid line), and $G_{bm}=0$, $g_{am}=0.1\omega_b$ (green dashed line). The other parameters are $\Delta_{c}=-1\omega_b$, $\kappa_a=\kappa_m=0.1\omega_b$, $\gamma_b=0.01\omega_b$, $\overline{n}_{a}=\overline{n}_{m}=0$, and $\overline{n}_{b}=0.2$.}\label{p-p weak entanglement}
\end{figure}

The beam-splitter interaction term $g_{am}(\delta m^{\dagger}\delta a+\delta m\delta a^{\dagger})$ in Eq. (\ref{Hlrint}) only represents the coherent population transfer between the magnetostatic mode $\delta m$ and the microwave mode $\delta a$, which cannot generate stationary entanglement between modes $\delta m$ and $\delta a$, while the parametric amplification term $G_{bm}(\delta m\delta b+\delta m^{\dagger}\delta b^{\dagger})$ describes the two-mode squeezing between the magnetostatic mode $\delta m$ and the deformation mode $\delta b$, which is the fundamental source of bipartite entanglement. As shown in Fig. \ref{p-p weak entanglement}, when one only considers the magnetostrictive interaction $G_{bm}$ between the mechanical and magnetostatic modes, and shuts the magnetic dipole interaction $g_{am}$ between the magnon and cavity modes, there is no entanglement between the microwave photon and the phonon (red line). Similarly, when one only selects the magnetic dipole interaction $g_{am}$, and neglects the magnetostrictive force $G_{bm}$, the steady-state photon-phonon entanglement is also absent (green line). However, the entanglement between the mechanical and cavity modes without direct interaction is obvious (blue line) when magnetostrictive effect and magnetic dipole interaction exist simultaneously, which means that bipartite entanglement can be transferred between different subsystems. First, when the magnetostatic mode is resonant with the Stokes sideband of the driving laser, i.e., $\Delta_m=-1\omega_b$, the magnon and phonon interact in the form of two-mode squeezing and then the entanglement between them can be obtained. Second, the resonant beam-splitter interaction between the magnon and the photon transfers the initial magnon-phonon entanglement to the subsystem composed of the photon and phonon under the $\Delta_{a}=\Delta_{m}$ condition. Thus, we obtain the steady-state photon-phonon entanglement $E_{N}$ near $\Delta_{m}/\omega_{b}=-1$ as shown in Fig. \ref{p-p weak entanglement}, which is comparable to the entanglement $E_{N}\simeq0.2$ achieved in the system considered by Li $et$ $al$.~\cite{Li:2018}.

\begin{figure}[!htbp]
\centering
{\includegraphics[width=1.00\linewidth]{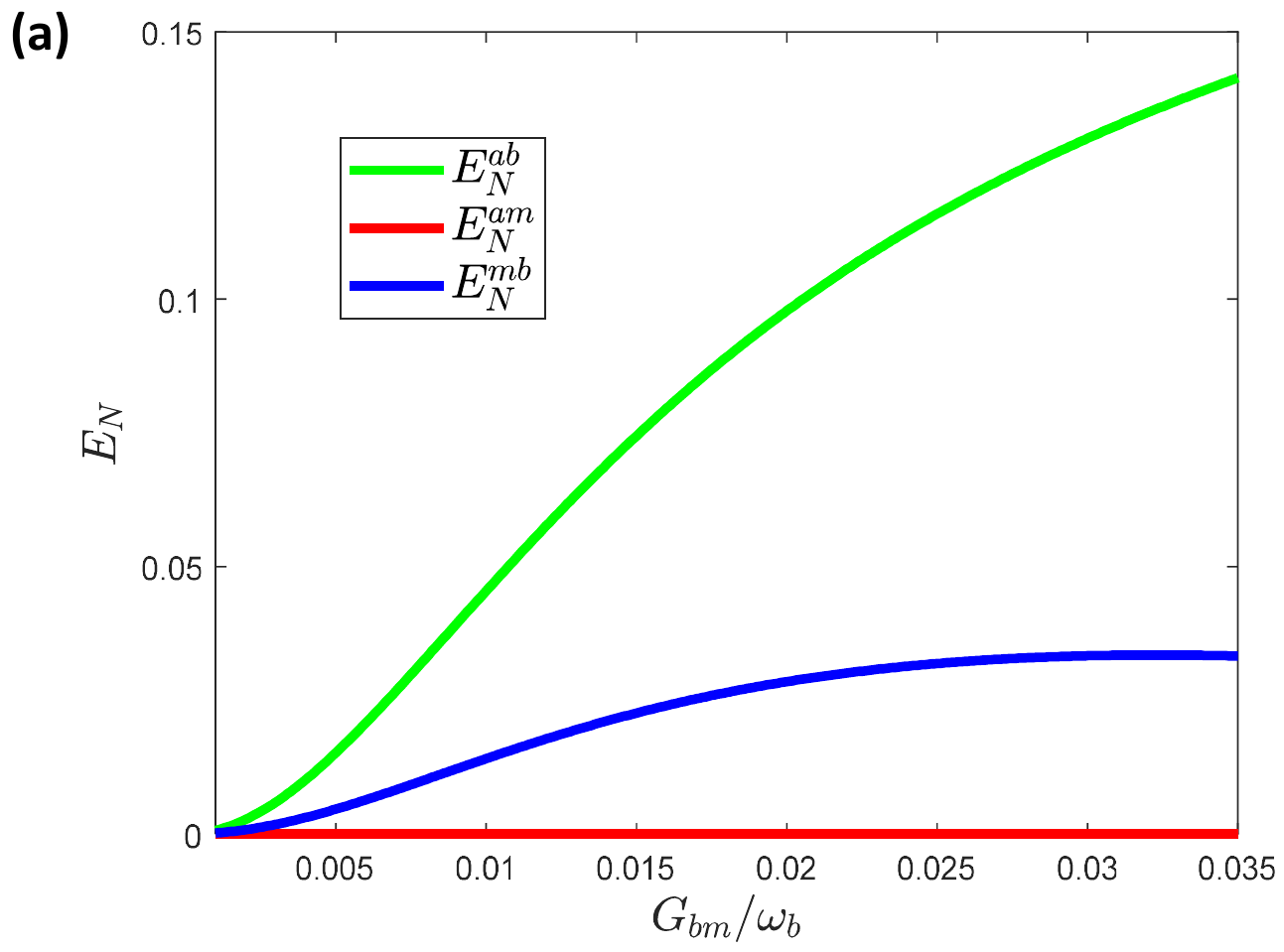}}\\
{\includegraphics[width=1.00\linewidth]{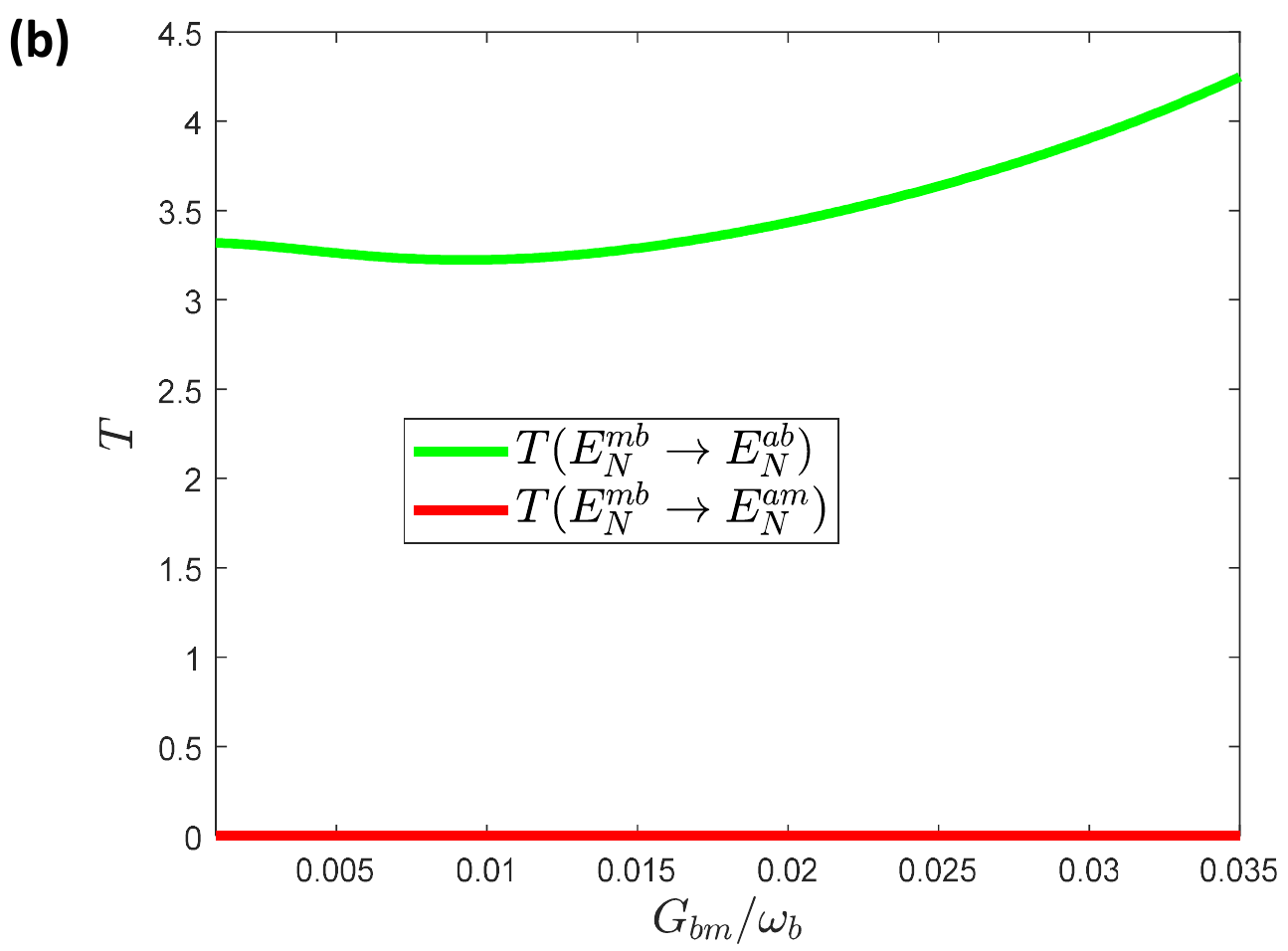}}
\caption{(Color online) (a) The stationary bipartite entanglement $E_N^{ab}$, $E_N^{am}$, and $E_N^{mb}$ and (b) the entanglement transfer rate $T$ versus the effective magnomechanical coupling $G_{bm}/\omega_b$ for $\Delta_m=-1\omega_b$ and $g_{am}=0.1\omega_b$. Other parameters are the same as the ones in Fig. \ref{p-p weak entanglement}.}
\label{entanglement transfer}
\end{figure}

To better understand the physics of entanglement transfer between different subsystems, we now define the transfer coefficient as follows
\begin{align}
T(E_N^{AB}\rightarrow E_N^{CD})=\frac{E_N^{CD}}{E_N^{AB}},\label{transfer rate}
\end{align}
where $A,B,C,D=m,a,b$. In Fig. \ref{entanglement transfer}$(a)$, we numerically show the steady-state bipartite entanglement $E_{N}^{ab}$ (photon-phonon), $E_{N}^{am}$ (photon-magnon), and $E_{N}^{mb}$ (magnon-phonon) versus the effective magnomechanical coupling $G_{bm}/\omega_{b}$ under the case of $g_{am}=0.1\omega_{b}$. We discover no entanglement between the photon and the magnon, i.e., $E_{N}^{am}=0$ (red line), while the magnon-phonon entanglement $E_{N}^{mb}$ and the photon-phonon entanglement $E_{N}^{ab}$ are observable, even the entanglement between the indirectly coupled cavity photon and phonon (green line) is larger than the entanglement between the directly coupled magnon and phonon (blue line), which is ascribed to the fact that the strong beam-splitter interaction $g_{am}$ between the photon and magnon instantaneously transfers from the magnon-phonon entanglement to the photon-phonon subsystem, and further suppresses the initial magnon-phonon entanglement. Furthermore, we demonstrate that the entanglement transfer rate $T(E_N^{mb}\rightarrow E_N^{ab})$ (green line) can be changed by adjusting the strength of the magnomechanical coupling $G_{bm}$, even making it greater than 1, as shown in Fig. \ref{entanglement transfer}$(b)$. This indicates that it is possible to find a set of parameters that allow the initial entanglement $E_{N}^{mb}$ to be fully transferred to the photon-phonon subspace and achieve the strong steady-state photon-phonon entanglement. This will be discussed in detail in next section.
\section{reservoir engineering strong quantum entanglement and its directional transfer}
The hybrid three-mode cavity magnomechanical system can be applied to realize the strong quantum entanglement between the cavity mode $\delta a$ and the deformation mode $\delta b$, we first introduce the delocalized Bogoliubov modes
\begin{align}
\beta_{1} &=\delta a \cosh r+\delta b^{\dagger} \sinh r, \nonumber\\
\beta_{2} &=\delta b \cosh r+\delta a^{\dagger} \sinh r,\label{Bogoliubov modes}
\end{align}
where $r=arc\tanh(G_{bm}/g_{am})$ is the two-mode squeezing parameter. Then the linearized bosonic Hamiltonian Eq. (\ref{Hlrint}) can be rewritten by using the delocalized Bogoliubov mode $\beta_{1}$ as
\begin{align}
H_{blv}=G(\delta m\beta_{1}^{\dagger}+\delta m^{\dagger}\beta_{1}),\label{Bogoliubov modes Hamiltonian}
\end{align}
where $G=\sqrt{g_{am}^{2}-G_{bm}^{2}}$ is the effective coupling strength between the magnon mode $\delta m$ and the delocalized mode $\beta_{1}$. We see that $\delta m$ and $\beta_{1}$ are coupled by the beam-splitter interaction, while the Bogoliubov mode $\beta_{2}$ is left entirely decoupled from the magnetostatic mode $\delta m$. Physically, this beam-splitter type of interaction Hamiltonian $H_{blv}$ cannot establish entanglement between the magnon mode $\delta m$ and the Bogoliubov mode $\beta_{1}$ as mentioned in Sec. III, but allows us to effectively realize laser cooling of the delocalized mode $\beta_{1}$ when the occupancy of the magnetostatic mode $\delta m$ is kept low. The joint ground state of the hybridized modes $\beta_{1}$ and $\beta_{2}$ is the two-mode squeezed vacuum state $|\psi_{s}\rangle=S(r)|0_{a},0_{b}\rangle$, where $S(r)=\exp[r(\delta a\delta b-\delta a^{\dagger}\delta b^{\dagger})]$ is the two-mode squeezing operator. The state $|\psi_{s}\rangle$ is the vacuum for both the Bogoliubov modes $\beta_{1}$ and $\beta_{2}$, so, when either of two modes $\beta_{1}$ or $\beta_{2}$ is cooled down to the ground state by the intermediate mode $\delta m$, we can obtain large stationary entanglement between the photon mode $\delta b$ and phonon mode $\delta b$. As a cold reservoir, the magnetostatic mode $\delta m$ must have a strong dissipation rate $\kappa_{m}$, which ensures that it can effectively cool the hybridized mode $\beta_{1}$ and is always in a low excited state satisfying the low excitation conditions as mentioned in Sec. II.

\begin{figure}[!htbp]
\centering
{\includegraphics[width=1.00\linewidth]{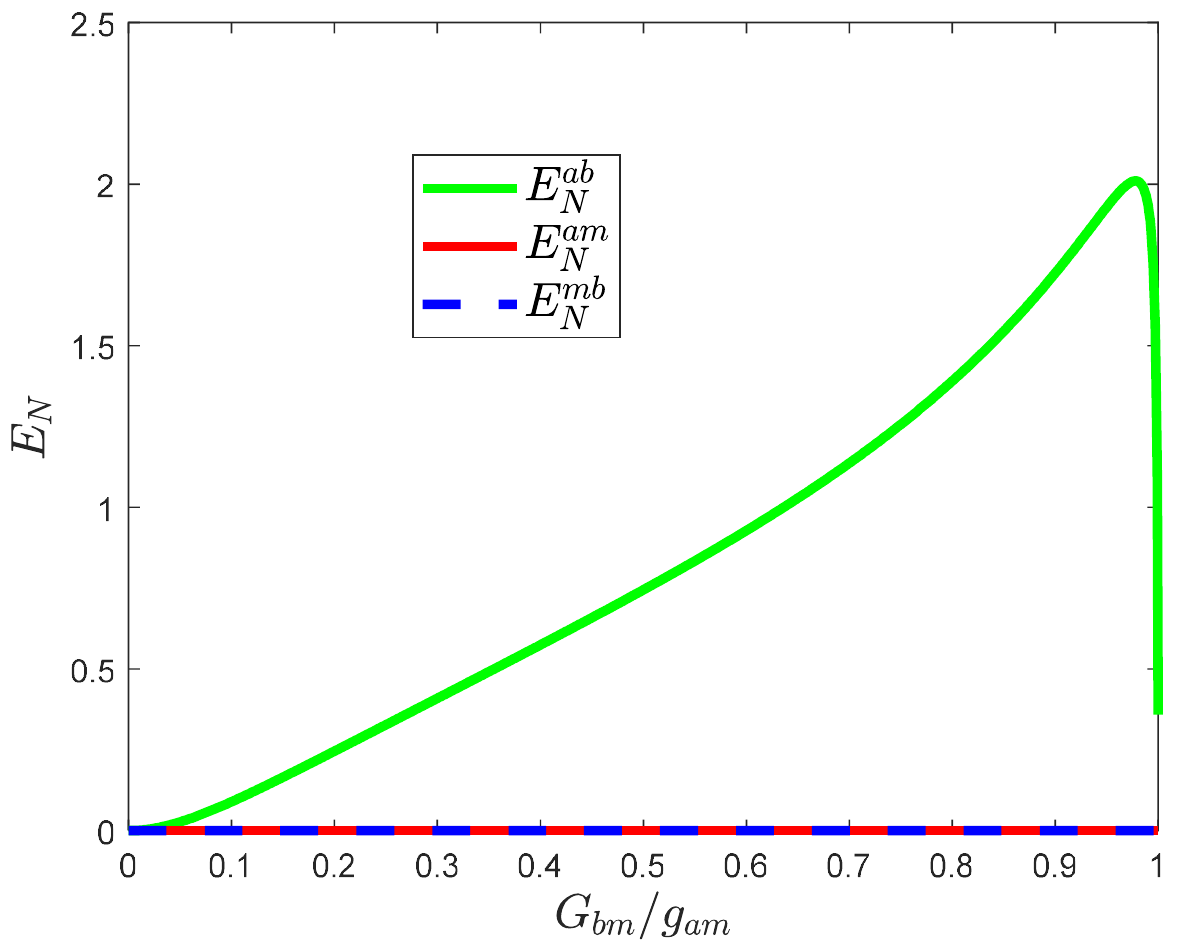}}
\caption{(Color online) The steady-state bipartite entanglement $E_N^{ab}$ (green solid line), $E_N^{am}$ (red solid line), and $E_N^{mb}$ (blue dashed line) in the three-mode cavity magnomechanical system as the function of the ratio $G_{bm}/g_{am}$. The other parameters are $g_{am}=0.65\omega_b$, $\Delta_c=\Delta_m=-1\omega_b$, $\kappa_m=0.1\omega_b$, $\kappa_a=\gamma_b=0.01\omega_b$, and $\overline{n}_{a}=\overline{n}_{m}=0$, $\overline{n}_{b}=0.2$.}\label{reservoir engineering entanglement}
\end{figure}

\begin{figure}[!htbp]
\centering
{\includegraphics[width=1.00\linewidth]{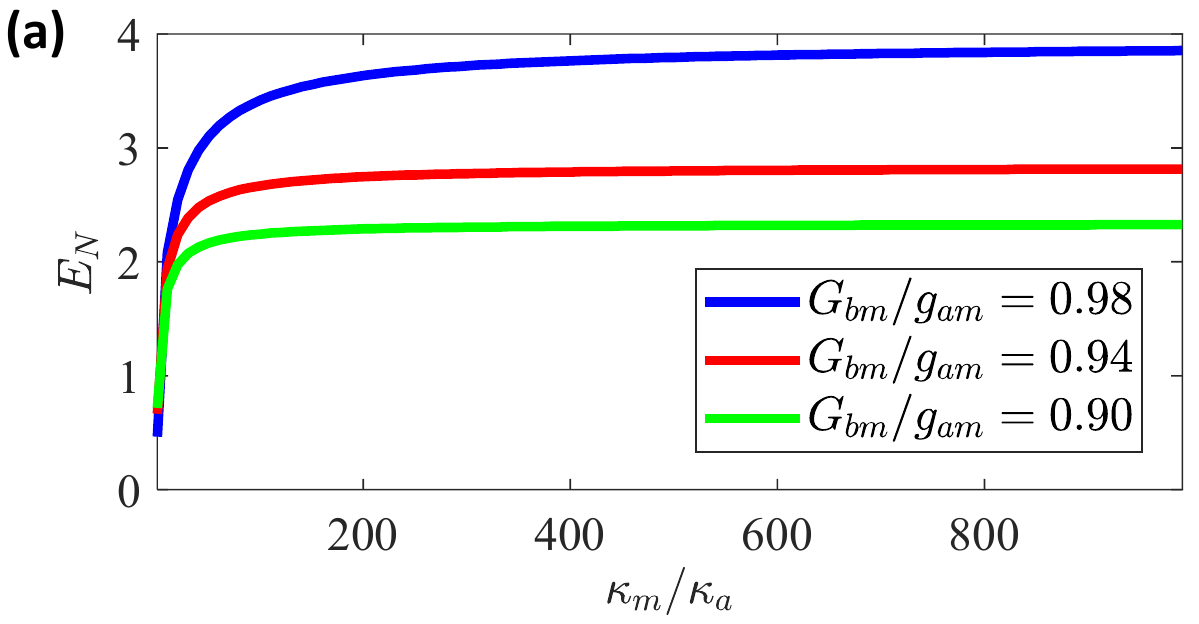}}\\
{\includegraphics[width=1.00\linewidth]{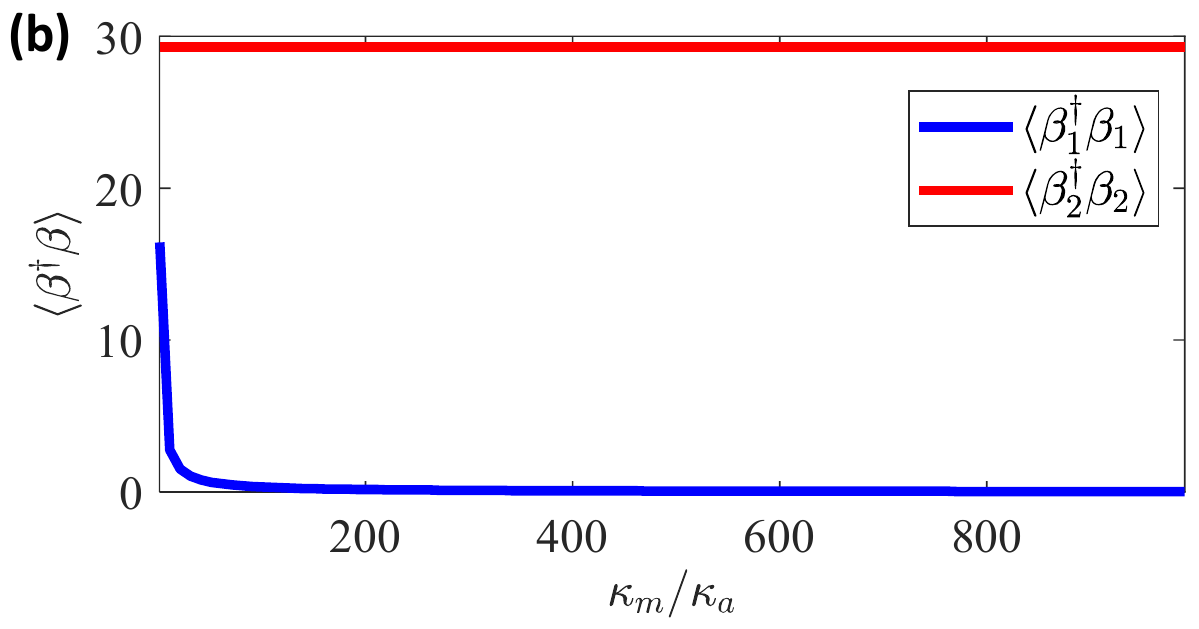}}
\caption{(Color online) (a) The steady-state photon-phonon entanglement $E_N$ versus the decay ratio $\kappa_m/\kappa_a$ for $G_{bm}/g_{am}=0.98$ (blue line), $G_{bm}/g_{am}=0.94$ (red line), and $G_{bm}/g_{am}=0.90$ (green line). (b) Steady-state occupancies of the Bogoliubov modes $\langle \beta_1^\dagger \beta_1\rangle$ (blue line) and $\langle \beta_2^\dagger \beta_2\rangle$ (red line) as functions of the decay ratio $\kappa_m/\kappa_a$ for $G_{bm}/g_{am}=0.98$. Other parameters are chosen the same as in Fig. \ref{reservoir engineering entanglement}.}
\label{decay entanglement}
\end{figure}

\begin{figure}[!htbp]
\centering
{\includegraphics[width=1.00\linewidth]{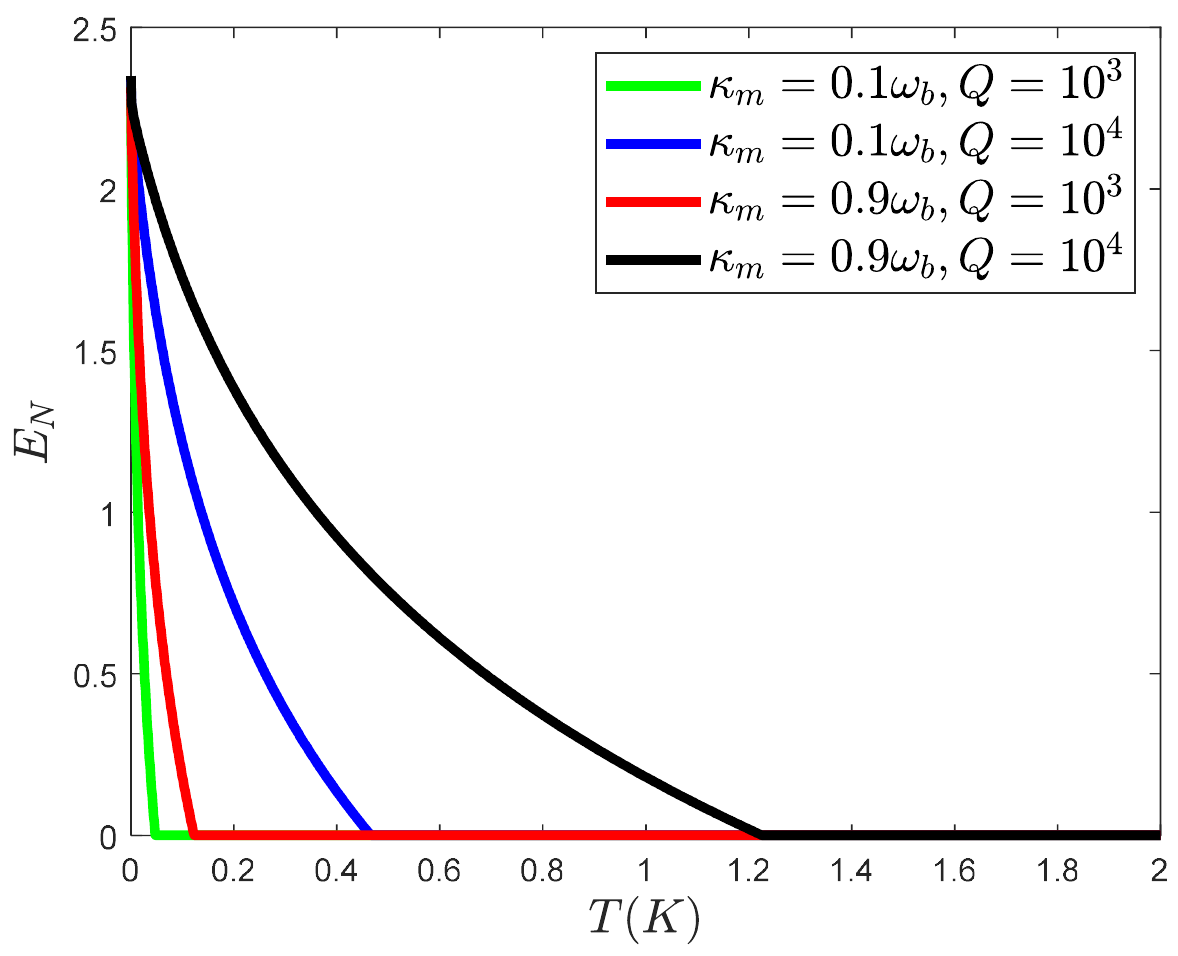}}\qquad
\caption{(Color online) The steady-state photon-phonon entanglement $E_N$ versus the temperature $T(K)$ of mechanical vibrator $\delta b$ for $\kappa_m=0.1\omega_b$, $Q=10^3$, $\kappa_m=0.1\omega_b$, $Q=10^4$, $\kappa_m=0.9\omega_b$, $Q=10^3$, and $\kappa_m=0.9\omega_b$, $Q=10^4$, respectively. Where $\kappa_a=\gamma_b$ and $G_{bm}=0.9g_{am}$, other parameters are the same as the ones in Fig. \ref{reservoir engineering entanglement}.}
\label{temperature entanglement}
\end{figure}

In Fig. \ref{reservoir engineering entanglement}, we show the steady-state entanglement $E_{N}$ between any two modes of three bosonic modes as a function of the effective coupling ratio $G_{bm}/g_{am}$. By selecting a set of reservoir-engineering parameters: $\kappa_m=0.1\omega_b$, $\kappa_a=\gamma_b=0.01\omega_b$, $\Delta_c=\Delta_m=-1\omega_b$, $g_{am}=0.65\omega_b$, one obtains the strong steady-state photon-phonon entanglement $E_{N}^{ab}$ (green solid line), which is enhanced roughly ten times compared with the scheme recently considered by Li $et$ $al$.~\cite{Li:2018}. The stationary entanglement $E_{N}^{ab}$ first monotonously increases as $G_{bm}/g_{am}$ and can saturate at values much larger than the maximum steady-state entanglement $E_{N}\sim \ln2$ achievable with a coherent two-mode squeezing interaction, as has been discussed in Refs.~\cite{WangYD:2013,Tian:2013}; then it will suddenly decrease to a smaller value as $G_{bm}/g_{am}\rightarrow1$, thanks to $r=arc\tanh(G_{bm}/g_{am})$ gradually increases upon the increase of the $G_{bm}$, the steady-state entanglement $E_{N}^{ab}$ is enhanced, but the thermal occupancies of these modes $\langle\beta_{j}^{\dagger}\beta_{j}\rangle$ are also increased, which in turn reduce the entanglement $E_{N}^{ab}$. However, the steady-state photon-magnon entanglement $E_{N}^{am}$ (red solid line) and the magnon-phonon entanglement $E_{N}^{mb}$ (blue dashed line) are almost always zero in the reservoir-engineering parameter regime. Furthermore, Fig. \ref{reservoir engineering entanglement} tells us that the entanglement transfer rate $T(E_N^{mb}\rightarrow E_N^{ab})\rightarrow\infty$. This indicates that the initial entanglement $E_{N}^{mb}$ is effectively transferred to the photon-phonon subspace and completely suppressed. In other words, one can achieve the entanglement directional transfer between different subsystems via the reservoir engineering approach.
In reservoir-engineering parameter regions, we have taken $\kappa_{m}=10\kappa_{a}=10\gamma_{b}$, increasing the dissipative difference between the cold reservoir mode $\delta m$ and the target modes $\delta a$ and $\delta b$ can theoretically further improve the steady-state photon-phonon entanglement. In Fig. \ref{decay entanglement}$(a)$, one shows the steady-state photon-phonon entanglement $E_{N}$ with $\kappa_{m}=0.1\omega_{b}$ versus the decay ratio $\kappa_m/\kappa_a$ for $G_{bm}/g_{am}=0.98$ (blue line), $G_{bm}/g_{am}=0.94$ (red line), and $G_{bm}/g_{am}=0.90$ (green line), respectively. We note that the steady-state entanglement $E_{N}$ can be significantly enhanced by increasing the dissipation ratio $\kappa_m/\kappa_a$. Moreover, by comparing these three curves in Fig. \ref{decay entanglement}$(a)$, we also show that the entanglement can be improved by adjusting the relative strength of the magnomechanical coupling $G_{bm}$ and the magnetic dipole interaction $g_{am}$. In Fig. \ref{decay entanglement}$(b)$, we numerically show that the steady-state occupancies of the Bogoliubov modes $\langle \beta_1^\dagger \beta_1\rangle$ and $\langle \beta_2^\dagger \beta_2\rangle$ under the case of $G_{bm}/g_{am}=0.98$ and $\kappa_{m}=0.1\omega_{b}$ as a function of the decay ratio $\kappa_m/\kappa_a$, where the occupancies of the two modes are explicitly given by
\begin{align}
\langle \beta_1^\dagger \beta_1\rangle=&\cosh^{2}r\langle\delta a^{\dagger}\delta a\rangle+\sinh^{2}r\langle\delta b^{\dagger}\delta b\rangle+\sinh^{2}r \nonumber\\
  &+\sinh r\cosh r\langle\delta a\delta b+\delta a^{\dagger}\delta b^{\dagger}\rangle,\label{occupancies1}
\end{align}
and
\begin{align}
\langle \beta_2^\dagger \beta_2\rangle=&\cosh^{2}r\langle\delta b^{\dagger}\delta b\rangle+\sinh^{2}r\langle\delta a^{\dagger}\delta a\rangle+\sinh^{2}r \nonumber\\
  &+\sinh r\cosh r\langle\delta a\delta b+\delta a^{\dagger}\delta b^{\dagger}\rangle.\label{occupancies2}
\end{align}
Since increasing the ratio $\kappa_m/\kappa_a$ enhances the ability of the reservoir mode $\delta m$ to cool the hybridized mode $\beta_{1}$, thus the Bogoliubov mode $\beta_{1}$ is rapidly cooled, corresponding to an abrupt reduction of the occupancy $\langle \beta_1^\dagger \beta_1\rangle$ (blue line). While the delocalized hybridized mode $\beta_{2}$ is decoupled from the magnon mode $\delta m$, so it cannot be cooled, i.e., the thermal occupancy $\langle \beta_2^\dagger \beta_2\rangle$ remains constant (red line).
Because the system will inevitably interact with the external environment, the steady-state entanglement will be affected by the thermal bath. In Fig. \ref{temperature entanglement}, we show the steady-state photon-phonon entanglement $E_{N}$ with $g_{am}=0.65\omega_{b}$ as a function of the temperature $T$ for different magnon mode dissipation rate $\kappa_{m}$ and different quality factor $Q=\omega_{b}/\gamma_{b}$ of the mechanical mode $\delta b$. For a magnetostatic mode with strong interaction with cold environment ($\kappa_{m}=0.9\omega_{b}$) and a mechanical mode with high quality factors $Q$, the death temperature $T$ of the steady-state entanglement $E_{N}$ can reach about $1.2K$, corresponding to the black line in Fig. \ref{temperature entanglement}. This is critical for detecting quantum entanglement experimentally. Next we will explore how to verify CV system entanglement in practical applications.

The steady-state photon-phonon entanglement defined by logarithmic negativity $E_{N}$ (see Appendix) can be verified by measuring all of the related entries of the corresponding covariance matrix~\cite{Vitali:2007,Liu:2021,Duan:2000,Paternostro:2007,DeChiara:2011,Li:2015}. The amplitude quadrature $\delta x_{1}$ and the phase quadrature $\delta y_{1}$ of the microwave cavity $\delta a$ can be directly measured by homodyning the cavity output. The position quadrature $\delta q$ and the momentum quadrature $\delta p$ of the mechanical deformation mode $\delta b$ can be measured by coupling with an additional optical cavity. The probe cavity will store information about the mechanical mode via the beam-splitter interaction, when one pumps the cavity with the driving at the red sideband. Then, the quadratures of the deformation mode can be read by observing the output field of the cavity. Finally, we can calculate the logarithmic negativity to get the degree of entanglement from all the terms of the covariance matrix that we have.

\section{Conclusion}

In conclusion, we have considered a hybrid magnomechanical setup with a microwave cavity and a YIG sphere, where the magnetostatic and mechanical deformation modes of YIG are coupled with each other by magnetostrictive interaction when YIG is driven directly by an electromagnetic field, simultaneously, the magnetostatic mode and the microwave cavity mode are coupled through magnetic dipole interaction. We have shown that the magnon-phonon entanglement induced by the two-mode squeezing interaction can be transferred to other subspaces by the beam-splitter interaction, so we obtain the stationary entanglement between the photon mode and the phonon mode that interact indirectly. Furthermore, one realizes the directional transfer of the bipartite entanglements and generates the strong steady-state photon-phonon entanglement via the reservoir engineering scheme, where the magnon mode is used for effectively laser cooling the Bogoliubov mode hybridized over the microwave cavity and the mechanical oscillator. Finally, our numerical results also indicate that the cooling capacity of the magnon mode can be further enhanced by simply increasing the ratio of the decay rate for the magnetostatic mode to that for the cavity mode, so that the entanglement is also improved, and the steady-state photon-phonon entanglement is robust against environmental temperature. The reservoir engineering scheme can generate some kind of specific bipartite entanglement according to the actual needs, and can potentially be generalized to other hybrid quantum systems based on magnonics, which will open new perspectives for quantum information processing.

\begin{acknowledgments}
This work was supported by National Natural Science Foundation of China (Grants No. 11874190 and No. 12047501). Support was also provided by Supercomputing Center of Lanzhou University.
\end{acknowledgments}

\appendix
\section{STEADY-STATE BIPARTITE ENTANGLEMENT}
Quantization and experimental measurement of Gaussian entanglement are essential to quantum information science. In this section, we will show how to describe quantum entanglement by mathematical language. First, one introduces the quadrature position operator and the quadrature momentum operator for three bosonic modes
\begin{align}
\delta x_{1}=\frac{\delta a+\delta a^{\dagger}}{\sqrt{2}} \, , \qquad
\delta y_{1}=\frac{\delta a-\delta a^{\dagger}}{i\sqrt{2}},\nonumber \\
\delta x_{2}=\frac{\delta m+\delta m^{\dagger}}{\sqrt{2}} \, , \qquad
\delta y_{2}=\frac{\delta m-\delta m^{\dagger}}{i\sqrt{2}},\nonumber \\
\delta q=\frac{\delta b+\delta b^{\dagger}}{\sqrt{2}} \, , \qquad
\delta p=\frac{\delta b-\delta b^{\dagger}}{i\sqrt{2}},
\end{align}
and their input quantum noises
\begin{align}
X_{1}^{in}=\frac{a_{in}+a_{in}^{\dagger}}{\sqrt{2}} \, , \qquad
Y_{1}^{in}=\frac{a_{in}-a_{in}^{\dagger}}{i\sqrt{2}},\nonumber \\
X_{2}^{in}=\frac{m_{in}+m_{in}^{\dagger}}{\sqrt{2}} \, , \qquad
Y_{2}^{in}=\frac{m_{in}-m_{in}^{\dagger}}{i\sqrt{2}},\nonumber \\
Q^{in}=\frac{b_{in}+b_{in}^{\dagger}}{\sqrt{2}} \, , \qquad
P^{in}=\frac{b_{in}-b_{in}^{\dagger}}{i\sqrt{2}}.
\end{align}
Assume a vector $M=[\delta x_{1},\delta y_{1},\delta x_{2},\delta y_{2},\delta q,\delta p]^{T}$, the linearized Langevin equations for the quantum fluctuations (Eq. (\ref{quantumEs})) can be written in compact form as
\begin{align}
\dot M=AM+N,\label{cform}
\end{align}
where
\begin{equation}
\begin{aligned}
A&=\begin{pmatrix}
-\frac{\kappa_{a}}{2} & \Delta_{c} & 0 & g_{am}& 0 & 0\\
-\Delta_{c} & -\frac{\kappa_{a}}{2} & -g_{am} & 0& 0& 0 \\
0 & g_{am} & -\frac{\kappa_{m}}{2} &\Delta_{m} & 0& 0\\
-g_{am} & 0 & -\Delta_{m} & -\frac{\kappa_{m}}{2}&-2 G_{bm}& 0\\
0 & 0 & 0 &0 & -\frac{\gamma_{b}}{2}& \omega_{b}\\
0 & 0 &-2 G_{bm} & 0&-\omega_{b}&-\frac{\gamma_{b}}{2}\\
\nonumber\end{pmatrix}\;,\label{matrix}
\end{aligned}
\end{equation}
with $N=[\sqrt{\kappa_{a}}X_{1}^{in},\sqrt{\kappa_{a}}Y_{1}^{in},\sqrt{\kappa_{m}}X_{2}^{in},
\sqrt{\kappa_{m}}Y_{2}^{in},\sqrt{\gamma_{b}}Q^{in},\\ \sqrt{\gamma_{b}}P^{in}]^{T}$ describes the thermal noise source of the system. According to the Routh-Hurwitz criterion~\cite{Vitali:2007,DeJesus:1987}, we know that the system is in a steady state when all of the eigenvalues of matrix $A$ have negative real parts. The input quantum noises of the system are Gaussian and the Hamiltonian is linearized, which ensures that the system is always in a Gaussian state, and the CV quantum entanglement of the system can be fully characterized by its $6\times6$ covariance matrix $\sigma$ with components defined as
\begin{align}
\sigma_{ij}=\langle M_{i}M_{j}+M_{j}M_{i}\rangle.\label{sigma}
\end{align}
When the stability conditions are satisfied, covariance matrix $\sigma$ will obey the following Lyapunov equation
\begin{align}
A\sigma+\sigma A^{T}=-D,\label{Lyapunov}
\end{align}
where $D$ is a diffusion matrix associated with the noise correlation functions and defined by
\begin{align}
D_{ij}\delta(t-t^{\prime})=\langle N_{i}(t)N_{j}(t^\prime)+N_{j}(t^\prime)N_{i}(t)\rangle/2,\label{Dfenl}
\end{align}
it is easy to verify that $D$ is diagonal
\begin{align}
D=&diag[\kappa_{a}(2\overline{n}_{a}+1),\kappa_{a}(2\overline{n}_{a}+1),\kappa_{m}(2\overline{n}_{m}+1),
\nonumber\\
&\times\kappa_{m}(2\overline{n}_{m}+1),
\gamma_{b}(2\overline{n}_{b}+1),\gamma_{b}(2\overline{n}_{b}+1)]/2.\label{Ddiag}
\end{align}
It is convenient to use the covariance matrix to calculate the bipartite entanglement between the cavity mode $\delta a$ and the deformation mode $\delta b$ by the reservoir engineering approach as in the main text. The reduced $4\times4$ covariance matrix $\sigma^{\prime}$ related to $\delta a$ and $\delta b$ can be extracted from the full $6\times6$ covariance matrix $\sigma$ by keeping the components in the $k$th rows and $l$th columns ($k,l\in\{1,2,5,6\}$), then we construct the two-mode covariance matrix $\sigma^{\prime}$ in the following form
\begin{align}
\sigma^{\prime}&=\begin{pmatrix}
R_{1} & R_{3} \\
R_{3}^{T} & R_{2} \\
\end{pmatrix}\;,\label{cm}
\end{align}
where $R_{1}$, $R_{2}$, and $R_{3}$ are $2\times2$ subblock matrices. The logarithmic negativity $E_{N}$ used to quantify the photon-phonon entanglement is then given by~\cite{Vitali:2007,Vidal:2002}
\begin{align}
E_{N}=\max[0,-\ln(2\eta)],\label{En}
\end{align}
where $\eta \equiv 2^{-1/2}\{\Sigma-[\Sigma^{2}-4\det\sigma^{\prime}]^{1/2}\}^{1/2}$, and $\Sigma=\det R_{1}+\det R_{2}-2\det R_{3}$.

\end{document}